\documentstyle [12pt] {article}
\begin{document}
\title {Stability of Stellar Objects Orbiting Sgr A*}
\maketitle
\begin{center}
{\bf { Magd E. Kahil{\footnote{ October University for Modern Sciences and Arts, Giza , Egypt \\  The American University in Cairo, New Cairo, Egypt \\
e.mail: kahil@aucegypt.edu}}} \footnote{Egyptian Relativity Group. Cairo, Egypt} }
\end{center}

\abstract{Path equations of different orbiting objects in the presence of very strong gravitational fields are essential to examine the impact of its  gravitational effect on the stability  of each system.
Implementing  an analogous method, used   to examine the stability of planetary systems by solving the geodesic deviation equations  to obtain a finite value of the magnitude of its corresponding  deviation  vectors.
Thus, in order to know whether a system is stable or not, the solution of corresponding deviation equations  may give an indication about the  status of the stability for orbiting systems.Accordingly, two questions must be addressed based on the status of stability of stellar objects orbiting super-massive black holes in the galactic center.\\
 1. Would the deviation equations play the same relevant role  of orbiting planetary systems for massive spinning objects such as neutron stars or black holes?\\
2. What type of field theory  which  describes  such  a strong gravitational field ?} \\

\\[2mm]
 \section{Introduction}
The problem of stability in our study is centered only on examining the stability of orbits in a very strong gravitational fields. In our Galaxy, S-stars are counted to be good candidates, to explain such a phenomenon. S-Stars are of spectral class B, that have been traced near infrared .  The characteristic behavior of these stars as they are very fast orbital motions around the Galactic Center, with orbital periods more than 16 years , high eccentricities $e > 0.2$, and their distances from the Galactic Center is between $ 10^0-10^2 mpc $ (Han (2014)), which  is greater than the center's radius $r >> r_{g}$ where $r_{g}$ its Schwarzschild radius. Also, a stringent condition is taken based on $\frac{m}{M} < 10^{-5}$, where $m$ and $M$ are masses of stellar object and center of SgrA* respectively (Iorio (2011)).
    One of  most brightest member of this group is S2, which takes about 16 years to revolve about the center of  Galaxy with a radial speed   $10,000km/sec$ and its mass is about $15 m_{sun}$ (Meyer et al (2012)).  Recently, another type of stars S0-102 with lesser brightness and shorter period about 11.5 years.
Unlike, the  orbits of satellites, planets or pulsars in the galactic center  the orbital periods are much longer leading to the relativistic effects increase more steeply with small radius and very high velocities than classical effects leading to the involvement of relativity is strongly appear around the peri-center passage. Accordingly, S-stars can be counted as clocks in orbit around a black hole moving on geodesics (Angelil et al.(2014)). Any slight effective perturbation on these trajectories can be obtained by obtaining its corresponding geodesic deviation equations.

 In  general the problem of stability is not only related to geodesic deviation equations ,but to  path deviation equations  of spinning object for a point mass particle [Mohseni 2010], which can also be extended to be charged and spinning charged objects. However, a slight problem can be emerged which is the solution of these deviation equations are completely affected by a coordinate system. Yet, Wanas and Bakry (2008) developed an approach based on determining a scalar value of the geodesic deviation capable for detecting the status of stability of any a certain planetary system in the presence of weak gravitational fields (Wanas and Bakry (2008)).

    In the present work, we are going to examine  stability conditions in the presence of a strong gravitational field, using Verozub's version of bi-metric  theory of gravity, which is one of the most appealing  theories (Verozub 2015). \\

\section{Equations of Motion for Orbiting Objects}
 It is well known that from observational methods , to confirm that both planetary and stellar objects are exhibiting  two types of motion revolving and spinning to become stable in their orbits. From this perspective, it is important to study stability of these systems by causing a slight perturbation that affects these combined motion and checks whether the object remains in the orbit or lose it forever. Such a technique is required to solve the path deviation equations of these objects. Accordingly, it is vital to obtain these equations from perturbing the original path equation. In case of planets/stellar objects, several authors have recommended Mathisson-Papapetrou- Dixon  equations (MPD) to be most reliable set of equation for describing such a situation (Dixon (1970)).
 \begin{equation}
\frac{DP^{\mu}}{DS}= F^{\mu},
 \end{equation}
 \begin{equation}
\frac{DS^{\mu \nu}}{DS}= M^{\mu \nu} ,
 \end{equation}
 where $P^{\mu}$ is the momentum vector, $ F^{\mu} = \frac{1}{2} R^{\mu}_{\nu \rho \delta} S^{\rho \delta} U^{\nu}$, $R^{\alpha}_{\beta \rho \sigma}$ is the Riemann curvature, $\frac{D}{Ds}$ is the covariant derivative with respect  to a parameter $S$,$S^{\alpha \beta}$ is the spin tensor, and $ M^{\mu \nu} =P^{\mu}U^{\nu}- P^{\nu}U^{\mu}$
 such that $U^{\alpha}= \frac{d x^{\alpha}}{ds}$ is the unit tangent vector to the geodesic. \\
 Using the following identity on both equations (1) and (2)
  \begin{equation}
  A^{\mu}_{; \nu \rho} - A^{\mu}_{; \rho \nu} = R^{\mu}_{\beta \nu \rho} A^{\beta},
  \end{equation}
  such that $A^{\mu}$ is an arbitrary vector.
 Multiplying both sides with  vectors, $U^{\rho} $ and $ \Psi^{\nu}$ as well as using the following condition (Heydri-Fard et al (2005))
 \begin{equation}
 U^{\alpha}_{; \rho} \Psi^{\rho} =  \Psi^{\alpha}_{; \rho } U^{\rho},
 \end{equation}
and $\Psi^{\alpha}$ is its deviation vector associated to the  unit vector tangent $U^{\alpha}$.
 Also in a similar way:

 $$
 S^{\alpha \beta}_{; \rho} \Psi^{\rho} =  \Phi^{\alpha \beta}_{; \rho } U^{\rho},
 $$

 one obtains the corresponding deviation equations (Mohseni (2010))
  \begin{equation}
\frac{D\Phi^{\mu}}{DS}= F^{\mu}_{; \rho} \Psi^{\rho},
 \end{equation}
 \begin{equation}
\frac{D\Phi^{\mu \nu}}{DS}= M^{\mu \nu}_{; \rho} \Phi^{\rho}
 \end{equation}
 where $\Phi^{\alpha}$, $\Phi^{\alpha \beta}$ are the spin path deviation and the spin tensor deviation associated to a path characterized by a parameter {S} and $(;)$ is the covariant derivative in Riemannian spaces. \\
  In our study, it is worth mentioning that in case of S-systems, the orbiting systems are becoming  MPD  with $S^{\mu \nu}$ is constant.\\
  Thus,
   \begin{equation}
\frac{DU^{\mu}}{DS}= \frac{1}{2m} F^{\mu},
 \end{equation}
 \begin{equation}
\frac{DS^{\mu \nu}}{DS}= 0 ,
 \end{equation}
 with taking into consideration that
 $$ S^{\mu \nu} U_{\mu} =0.$$
  Accordingly, one transform $V^{\alpha}$ to $U^{\alpha}$  in the following way:
$$
   V^{\alpha} = U^{\alpha} + \sigma \frac{D \Psi^{\alpha}}{Ds}
$$
where $V^{\alpha}= \frac{d x }{d S}$ the tangent vector describing the spinning motion, ${S}$ its associated geodesic parameter and $\sigma$ ia an arbitrary parameter acting as a spin angular momentum ratio (Bini and Gerlalico (2014)).\\
Thus,
 $$
  \frac{D}{D S} V^{\alpha} = \frac{D}{D s}( U^{\alpha} + \sigma \frac{D \Psi^{\alpha}}{Ds}) \frac{ds}{d S}
 $$
as well as
 $$
 S^{\mu \nu} = {\hat{s}}( \Psi^{\mu}U^{\nu} - \Psi^{\nu}U^{\mu}) ,
 $$
 such that  $ \sigma = \frac{\hat{s}}{m}$.
 Thus,
 $$
  \frac{D}{D S} V^{\alpha} = \frac{\hat{s}}{m}R^{\mu}_{\nu \rho \sigma} U^{\rho}\Psi^{\sigma} U^{\beta} \frac{ds}{dS},
 $$
 Let $\frac{ds}{dS} =1$,
 we  obtain
  \begin{equation}
  \frac{D}{D S} V^{\alpha} = \frac{\hat{s}}{m} R^{\mu}_{\nu \rho \sigma} U^{\rho} \Psi^{\sigma} U^{\beta} ,
  \end{equation}
  the above equation gives an indication that the path equation of a spinning particle is expressed in terms of its corresponding geodesic deviation vector. \\

 Such a result can be also extended to study the  motion of binary pulsar ,PSR-J0737-3039. It is  composed of two neutrons  stars, located at a distance $10^9 km$ from the Galactic Center, of negligible intrinsic rotations regrading to the orbital  period of about 2.4 hours , and their total mass is about $0.7M_{sun}$ . This can give that
 $$
  \frac{D}{D S} V_{1}^{\alpha}- \frac{D}{D {S}}V_{2}^{\alpha} = \frac{D}{D s}( U^{\alpha} +
   \sigma_{1} \frac{D \Psi^{\alpha}}{Ds}) \frac{ds}{d {S}}
 $$
$$
   ~~~~~~~~~~~~~~~~~~~~~~~~~~~~~~~~- \frac{D}{D s}( U^{\alpha} + \sigma_{2} \frac{D \Psi^{\alpha}}{Ds}) \frac{ds}{d {S}}.
$$
 where , $ \sigma_{1}$ and $\sigma_{2}$ the angular momentum ratio of each neutron star of this binary pulsar, to obtain the following equation

 $$
 \frac{D}{D {S_{1}}} V_{1}^{\alpha} - \frac{D}{D{S_{2}}} V_{2}^{\alpha}  =  ( \sigma_{1} - \sigma_{2}) R^{\alpha}_{\rho \delta \beta} U^{\delta} U^{\rho} \Psi^{\beta},
 $$
 in which  $V^{\alpha}_{1}$ and $V^{\alpha}_{2}$ are two  tangent vector associated to each spinning object in the binary system.\\
 Consequently, we find out that
\begin{equation}
\frac{D \bar{V^{\alpha}}}{DS} = ( \sigma_{1} - \sigma_{2}) R^{\alpha}_{\rho \delta \beta} U^{\delta}U^{\rho}\Psi^{\beta}
\end{equation}
such that,
$\bar V^{\alpha} = V^{\alpha}_{1}- V^{\alpha}_{2}.$  \\

 \section{The Relationship between Stability and Geodesic Deviation}
 It is well known that stability of planetary {/} stellar systems can be represented by a path deviation equation for an orbiting object.\\
 Consequently,  the stability tensor can be defined as follows:
\begin{equation}
 H^{\alpha}_{\gamma} \Psi^{\gamma}= R^{\alpha}_{\beta \omega \gamma} U^{\beta}U^{\omega}
 \end{equation}
 where , $H^{\alpha}_{\gamma}$  is the stability tensor defined as (Di Bari and Cipriani (2000)).

 Thus geodesic deviation equation may be expressed in terms of stability tensor;
 \begin{equation}
\frac{D^{2} \Psi^{\alpha}}{DS^{2}} = H^{\alpha}_{\beta} \Psi^{\beta}
 \end{equation}
 which is reduced to
 \begin{equation}
   \frac{d^{2}\Psi^{\mu} }{dS^2} + 2 \Gamma^{\mu}_{\nu \rho} \frac{d \Psi^{\nu}}{dS} U^{\rho} + \Gamma^{\mu}_{\nu \rho , \sigma} U^{\nu} U^{\rho} \Psi^{\sigma} = 0,
   \end{equation}
provided that (Di Bari and Cipriani (2000))
   $$
   g_{\mu \nu} \Psi^{\mu} \Psi^{\nu} = constant.
   $$
   Also, equation (9) and (10) can be written in terms of stability tensor in following way
   \begin{equation}
  \frac{D}{D S} V^{\alpha} = \sigma R^{\mu}_{\sigma} \Psi^{\sigma} ,
  \end{equation}
   and
   \begin{equation}
\frac{D \bar{V^{\alpha}}}{DS} = ( \sigma_{1} - \sigma_{2}) H^{\alpha}_{\beta} \Psi^{\beta}.
\end{equation}
Such a result for linearized systems gives rise to indicate that geodesic deviation vector can determine the spin path equation for S-stars and binary pulsar that are expressed by MPD equations.
   In order to obtain the solution, one must solve its corresponding field equation and define a certain coordinate system ,to obtain the value of the deviation vector. \\
    However, Wanas and Bakry (1995) introduced an approach , for examining the stability problem for any planetary system,  being a covariant coordinate independent which can be explained in the following way (Wanas and Bakry 1995)). \\ Let $\Psi^{\alpha}(S)$ is obtained from the solutions of the deviation equation in a given interval [a,b] in which $\Psi^{\alpha}(S)$ behave monotonically. These quantities can become sensors for measuring the stability of the system are

   \begin{equation}
 q^{\alpha}~~{\stackrel{def.}{=}}~~\Psi^{\alpha}(S)= C^{\alpha}f(S),
   \end{equation}
where $ C^{\alpha}$ are constants and $f(S)$ is a function known from the metric. If $ f(S) \rightarrow \infty$ , the system becomes unstable otherwise it is stable. This approach has been applied previously in examining the stability of some cosmological models (Wanas and Bakry (1995)) using two geometric structures (Wanas (1986)). The above approach has been modified by obtaining the scalar value of the deviation vector which gives rise to become independent of any coordinate system (Wanas and Bakry (2008))\\

    \begin{equation}  q~~ {\stackrel{def.}{=}}~~ \lim_{s \rightarrow b} \sqrt{\Psi^{\alpha}\Psi_{\alpha}} . \end{equation}   If  $q \rightarrow \infty$  then the system is unstable, otherwise it is always stable.

   Now for spinning objects with precession, we suggest  the above condition be extended to include the spin deviation tensor $\Phi^{\mu \nu}$ as
   \begin{equation}
    \bar{q}~~ {\stackrel{def.}{=}}~~ \lim_{s \rightarrow b}\sqrt{\Phi^{\alpha \beta}\Phi_{\alpha \beta}}.
    \end{equation}
    Thus, for such a member in stellar{/}planetary system is stable, if and only if  the magnitude of the scalar value of both spin deviation vectors ${\Phi}^{\alpha}$  and spin deviation tensors $\Phi^{\alpha \beta}$  to be real numbers respectively.
   i.e. either $ q \rightarrow \infty$ or $\bar{q} \rightarrow \infty $ the assigned member is unstable.
 Accordingly,  a strong stability condition must be admitted if both $q$ and $   \bar q$ are satisfying the following conditions :
\begin{equation}
\lim_{s \rightarrow \infty} (\Phi_{\alpha}\Phi^{\alpha}) =0,
\end{equation}
and
 \begin{equation} \lim_{s \rightarrow \infty} (\Phi_{\alpha \beta}\Phi^{\alpha \beta}) =0. \end{equation}
\\ \\
\section{Geodesic and Geodesic Deviation: The Bazanski Approach}
Geodesic and geodesic  deviation equations can be obtained simultaneously by using the Bazanski Lagrangian ((Bazanski (1989)):
\begin{equation}
L= g_{\alpha \beta} U^{\alpha} \frac{D \Psi^{\beta}}{DS},
\end{equation}
where $L$ is the lagrangian function. \\
Thus, it can be found clearly, if one takes the variation with respect to the deviation vector ${\Psi^{\rho}}$ to get geodesic equations:
\begin{equation}
\frac{dU^{\alpha}}{dS} + \Gamma^{\alpha}_{\mu \nu}U^{\mu}U^{\nu}=0.
\end{equation}
Also, the same technique can be applied to get the variation with respect to the tangent vector $U^{\rho}$ to get the geodesic deviation equations:
\begin{equation}
\frac{D^2 \Psi^{\alpha}}{DS^{2}} = R^{\alpha}_{. \beta \gamma \delta} \Psi^{\gamma} U^{\beta}U^{\delta}.
\end{equation}
The above Lagrangian has been modified to describe the path equation of a charged object to take the following form (Kahil (2006));
$$
L = g_{\alpha \beta} U^{\alpha}\frac{D \Psi^{\beta}}{DS} + \frac{e}{m} F_{\alpha \beta}U^{\alpha} \Psi^{\beta}
$$
 where $\frac{e}{m}$ is the ratio of charge to mass of any charged object, $F_{\mu \nu}$ is an electromagnetic field tensor. Taking the variation with respect to $\Psi^{alpha}$ one obtains
\begin{equation}
\frac{dU^{\alpha}}{d\bar{S}} +{\Gamma}^{\alpha}_{\mu\nu}U^{\mu}U^{\nu}= \frac{e}{m}F^{\mu}_{. \nu} U^{\nu}.
\end{equation}
While taking the variation with respect to $U^{\alpha}$  one obtains its corresponding deviation equations :
\begin{equation}
\frac{D^{2}\Psi^{\alpha}}{DS^{2}}= R^{\alpha}_{.\mu \nu\rho}U^{\mu}U^{\nu}\Psi^{\rho} +\frac{e}{m}(F^{\alpha}_{.\nu} \frac{D \Psi^{\nu}}{Ds}+F^{\alpha}_{. \nu ; \rho}U^{\nu}\Psi^{\rho})
\end{equation}

Also the corresponding Papapetrou Equation for rotating objects without precession can be obtained from the following Lagrangian:

\begin{equation}
L = g_{\alpha \beta} U^{\alpha}\frac{D \Psi^{\beta}}{DS} + \frac{1}{2m}F_{\mu}\Psi^{\mu}
\end{equation}

Taking the variation with respect to $\Psi^{\alpha}$, we obtain the spin path equation,
\begin{equation}
\frac{dU^{\alpha}}{dS}+\Gamma^{\alpha}{\mu \nu} U^{\mu}U^{\nu}= \frac{1}{2} F^{\alpha}
\end{equation}
and taking the variation with respect to $U^{\alpha}$, we obtain the spin deviation equation
\begin{equation}
\frac{D^{2}\Psi^{\alpha}}{DS^{2}}= R^{\alpha}_{\beta \gamma \delta}U^{\beta} U^{\gamma} \Psi^{\delta} + \frac{1}{2} F^{\alpha}_{; \rho}\Psi^{\rho}
\end{equation} .
 In case of the Dixon equation for spinning charged objects can be obtained in a similar way from the following Lagrangian

 \begin{equation}
L = g_{\alpha \beta} U^{\alpha}\frac{D \Psi^{\beta}}{DS} + \frac{1}{2m}(F_{\mu}+ e F_{\mu \nu} U^{\nu})\Psi^{\mu} .
\end{equation}
Taking the variation with respect to ${\Psi^{\mu}}$ we obtain

\begin{equation}
\frac{dV^{\alpha}}{d S}+\Gamma^{\alpha}_{\mu \nu}U^{\mu}U^{\nu}=   \frac{e}{m}F^{\mu}_{. \nu} U^{\nu}+\frac{1}{2m} F^{\mu}.
\end{equation}
While its corresponding deviation equation can be obtained by taking the variation with respect to $U^{\alpha}$
\begin{equation}
\frac{D^{2}\Psi^{\alpha}}{DS^{2}}= R^{\alpha}_{.\mu \nu\rho}U^{\mu}U^{\nu}\Psi^{\rho} +\frac{e}{m}{(F^{\alpha}_{.\nu} U^{\nu})}_{; \rho} \Psi^{\rho} + \frac{1}{2} F^{\alpha}_{; \rho}\Psi^{\rho}.
\end{equation}

  Similarly, we can modified the Lagrangian [21] to obtain spin equation and  spin deviation equation for rotating objects with precession in the following way:
\begin{equation}
L= g_{\alpha \beta} P^{\alpha} \frac{D \Psi^{\beta}}{DS} + F_{\alpha}\Phi^{\alpha}
\end{equation}
where $$ P^{\alpha}= m U^{\alpha} + U_{\beta}\frac{D S^{\alpha \beta}}{D S}.$$
In order to obtain an equation of spinning object with precession,  we  take the variation with respect to the deviation vector $\Phi^{\alpha}$
\begin{equation}
 \frac{D P^{\alpha}}{DS}= F^{\alpha}.
\end{equation}
And for its spin deviation equation, we take the variation with respect to $U^{\alpha}$ to become:
\begin{equation}
\frac{D^{2}\Phi^{\alpha}}{DS^{2}}= R^{\alpha}_{.\mu \nu\rho}P^{\mu}U^{\nu}\Phi^{\rho} + F^{\mu}_{; \rho} \Phi^{\rho}.
\end{equation}
While for its precession part it can be obtained using the following condition:
$$
  P_{\mu}S^{\mu \nu} =0 ,
$$
to give
$$
\frac{D S^{\alpha \beta}}{DS} = P^{\mu}U^{\nu}- P^{\nu}U^{\mu}.
$$

\section{Stability of Motion in Bimetric Theory of Gravity: The Verozub Approach}
In this section, we are showing that the treatment of the stability problem  in strong fields may be explained in the presence of bimetric theory of gravity.
  This type of bimetric theories was proposed by Rosen in 1940, who regarded  gravity can be expressed in flat space. Due to considering that, all objects of the Riemannaian space are functions in Minkowski space (Rosen 1973). But such a type of visualization gives no physical meaning, with inconsistency with observations as well as there is no relation between the two metrics (Verozub 2015).
     Recently, Verozub has introduced a new version of bimetric theory of gravity, stemmed from  a well known principle of Poincare that properties of space-time are relative to the properties of used measuring instruments,together with the Einstein idea of the relativity of properties  of space-time with respect to the distribution of matter  (Verozub 2008 ).

     It is well known that in general relativity that test particles in gravitational field move on geodesics in a Riemanannian space. Accordingly, one may figure out that the differential equations for obtaining the metric tensor $g_{\mu \nu}(x)$ of any distribution of matter must keep the geodesic equations invariant under coordinate transformations. Surprisingly, it can be found that these equations are also invariant under geodesic mapping of space time $V$ into $\bar V$ upon replacing $\Gamma^{\mu}_{\alpha \beta} \rightarrow \bar{\Gamma}^{\mu}_{\alpha \beta}$ of the Christoffel symbols in any fixed coordinate system to become
     \begin{equation}
     \bar{\Gamma}^{\mu}_{\alpha \beta} = {\Gamma^{\mu}_{\alpha \beta}} + \delta^{\mu}_{\alpha} \phi_{\beta} + \delta^{\mu}_{\beta} \phi_{\alpha}
     \end{equation}
  where $\phi_{\mu}(x)$ is a vector field. Moreover, transformations of the metric tensors are obtained by solving the partial differential equation
  \begin{equation}
  \bar{g}_{\mu \nu ; \alpha} = 2 \phi_{\alpha}(x)\bar{g}_{\beta \gamma}(x)+\phi_{\beta}(x)\bar{g}_{\gamma \alpha}(x)+ \phi_{\gamma}(x)\bar{g}_{\alpha \beta}(x),
  \end{equation}
  in which the semi-colon is related here to the covariant derivative of $V$.

  Thus, this field can be expressed in a Riemannian  space in terms of two metrics before and after the geodesic mapping of from one space time into another in the following way:
  \begin{equation}
     \phi_{\alpha} = \frac{1}{n+1}( \bar{\Gamma}^{\mu}_{\alpha \mu} - {\Gamma^{\mu}_{\alpha \mu}} )= \frac{1}{2(n+1)}{\frac{\partial}{\partial   x^{\alpha}}}ln|\frac{\bar{g}}{g}|.
   \end{equation}

     Thus, Verozub's version of bimetric  theory of gravity has two important results, geodesic transformations are playing the role of gauge transformations while coordinates transformation are acting the same way as in electrodynamics. It also gives a full description of motion of small particles of a perfect isentropic fluid able to describe gravity in strong gradational fields of a super-massive black hole Sgr A* at the Galactic Center. Also, the theory has neither singularities nor event horizon.

     From this perspective, we aim to study stability of orbiting objects like S2 and binary pulsar PSR-J0737-3039, by obtaining  their geodesic and geodesic deviations vectors. \\
         Implementing Verzob's version one can find that the trajectories of a test particles are   geodesics are  in  the co-moving reference frame, (CRF), described by $g_{\mu \nu}(\psi)$, such that $\psi_{\mu \nu}$ is a tensor field of spin 2  gravity, as found in Riemananain space of non zero curvature .  While, the same test particle is observed in an inertial reference frame (IRF)  as a point mass  moves  under the influence of a force field $\psi_{\mu \nu}$, as existed in  Minkwoskian space (Verozub 2008).

   Accordingly, the line element of the IRF is defined as follows:
  \begin{equation}
   d{\sigma}^2 =  \eta_{\mu \nu}(x) dx^{\mu} dx^{\nu}
   \end{equation}
  where $\eta_{\mu \nu}$ is the Minkowski metric and its corresponding CRF line element is defined as
   \begin{equation}
   dS^2 = g_{\mu \nu} (\psi) dx^{\mu}dx^{\nu}
   \end{equation}
   leading to define its corresponding affine connection:
   $$
   \bar\Gamma^{\alpha}_{\beta \rho} = \frac{1}{2} g^{\alpha \delta}(\psi) ( g_{\beta \delta, \rho} + g_{ \delta \rho, \beta} - g_{\beta \rho , \delta}).
$$
 Applying the Bazanski approach, we obtain  geodesic and geodesic deviation equations  of Verozub's version for bimetric theory of gravity:
\begin{equation}
L(\psi)= g_{\alpha \beta}(\psi) U^{\alpha} \frac{D \Psi^{\beta}}{DS}
\end{equation}
This can be seen clearly if one takes the variation with respect to the deviation vector ${\Psi^{\rho}}$ to get the geodesic equations:
\begin{equation}
\frac{dU^{\alpha}}{dS} + \bar\Gamma^{\alpha}_{\mu \nu}(\psi)U^{\mu}U^{\nu}=0
\end{equation}
Also, the same technique can be applied to get the variation with respect to the tangent vector $U^{\rho}$ to get the geodesic deviation equations:
\begin{equation}
\frac{D^{2}\Psi^{\alpha}}{DS^{2}} = \bar{R}^{\alpha}_{. \beta \gamma \delta}(\Psi) \Psi^{\gamma} U^{\beta}U^{\delta}
\end{equation}
where $\bar R$ is the Riemann Curvature described by the affine connection $\bar\Gamma^{\alpha}_{\beta. \rho \sigma}$ for the (CRF).
Thus the stability equation in this case becomes:
\begin{equation}
\frac{D^{2}\Psi^{\alpha}}{DS^{2}} = \hat{H}^{\alpha}_{.\gamma}(\psi) \Psi^{\gamma}.
\end{equation}
as $ \hat H^{\alpha}_{. \beta}$ is the stability tensor defined in CRF.
 Thus, the deviation vector  in CRF in Riemannian space can be expressed as a  separation vector of these particles under the action of a force field $\psi_{\mu \nu}$  in Minkowkian space ,which can be reduced to (Verozub (2015)) :
 \begin{equation}
\frac{\partial^2 \eta^{\alpha}}{\partial^{2} \tau} + \frac{\partial^{2}U}{\partial x^{\alpha}\partial x^{\beta} } \eta^{\beta} =0,
\end{equation}
where  $ \eta^{\alpha} = \frac{\partial x^{\alpha}}{\partial \xi}$ ,$x^{\mu} =x^{\mu}(\tau, \xi)$ such that $\eta^{\alpha}$ is  the separation vector and $U$ is the gravitational potential as measured in the flat space.
If we apply the Wanas-Bakry condition on the scalar of the separation vector between two geodesics  in a Minkowski space we can easily find
$$ \tilde{q} = \lim_{\tau \rightarrow \bar{b} } \sqrt{\eta^{\alpha}\eta_{\alpha}} $$ , where the solutions of ${\eta (\tau})$ in a given interval $[\bar{a}, \bar{b}]$ behave monotonically . If $\tilde{q} \rightarrow \infty$  then the system is unstable,
otherwise it is always stable.
Consequently, the strong stability condition becomes
\begin{equation}
\lim_{\tau \rightarrow \infty} (\eta_{\alpha}\eta^{\alpha}) =0.
\end{equation}
 Accordingly, we can conclude  that in a strong gravitational field, in covariant stability condition is examined by obtaining the scalar value of its associated separation vectors as defined in IRF rather than its equations. Such an approach gives the finiteness of the scalar value for the separation vector an indicator to decide whether the orbiting system is stable or not.

\section{Discussion and Concluding Remarks}
In this study,we have examined the stability of rotating objects in the presence of very strong gravitational field. One of most promising theories is the bimetric version of Verzub. The objects are considered as test particles due to to the stringent  condition $\frac{m}{M} < 10^{-5}$, e.g the S-stars are considered as test particles moving on geodesics and acting as clocks for the SgrA*.
  It has been assumed that the stability criterion may be estimated its status by extending the covariant stability condition method of Wanas-Bakry to examine S-stars and PSR J0737-3039. The stability of these systems are mainly dependent on obtaining the corresponding deviation vectors and then finding their scalar value in each case. Yet, an additive step may be obtained due to Veozub's bimetric theory, is the scalar part of the separation vectors obtained in IRF as defined in flat space is becoming a good candidate to examine the stability condition.

 Moreover, we have obtained a relationship between the spin tensor of a rotating object with its corresponding deviation vector. This result leads to identify the stability condition without finding out the spin deviation vector as an indicator of stability conditions, and examining only the stability condition on their corresponding deviation vector. Accordingly, we have obtained  a quick method to estimate whether the system is stable or not without going to lengthy calculation to determine the scalar value for the spin deviation vector, such an advantage works in favor of testing stability conditions for S-stars or binary pulsars orbiting SgrA*.    \\ \\
{\bf Acknowledgments}\\ [2mm] The author would like to thank Professor A. Zhuk and the organizing committee of the fifth Gammov Conference of astronomy , cosmology, astrophysics and  astrobiology in Odessa, Ukraine for their hospitality and good atmosphere of discussions during the conference period.
\\[3mm]
%\indent
\section*{\bf References}
Han, Wen-Biao (2014) Astrophysics and Astronomy {\bf{14}}, 1415 \\
Iorio, L. (2011) Phys. Rev. {\bf{D84}},124001. \\
Angelil, R. and Saha, P. (2014) MNRS,{\bf 444(2)}, 3780. \\
Meyer et al. (2012) Science {\bf{125506}}. \\
Wanas, M.I. and Bakry, M.A. (2008) Proc. MGXI part C, 2131. \\
Wanas, M.I. and Bakry, M.A. (1995), Astrophys. Space Sci. {\bf{228}}, 239. \\
Wanas, M.I. (1986) Astrophys. Space Sci. {\bf{127}}, 21. \\
Heydrai-Fard, Mohseni, M. and Sepanigi, H.R. (2005) Physics letters B, {\bf{626}}, 230. \\
Mohseni, M. (2010) Gen. Rel. Grav. {\bf{42}}, 2477.\\
Verzub, L.  (2015) {\it{Space-time Relativity and Gravitation}}, Lamberg Acadamic Publishing.  \\
Bini, D. and Geralico, A. (2014) Phys Rev.,{\bf{D84}},104012. \\
Bazanski, S.L. (1989) J. Math. Phys., {\bf {30}}, 1018. \\
Dixon, W. G. (1970)  Proc. R. Soc. London, Ser. A {\bf{314}}, 499 \\
Di Bari, Maria and Cipriani, P. (2000) Chaotic Universe, 444. \\
Papapetrou, A. (1951), Proceedings of Royal Society London A {\bf{209}} , 248  \\
Kahil, M.E. (2006), J. Math. Physics {\bf {47}},052501. \\
Rosen, N. (1973)  Gen. Relativ.  and Gravit., {\bf{4}},  435. \\
Verozub, L. (2008) Annalen der Physik, {\bf{27}}, 28. \\

\end{document}